\documentclass[prl,twocolumn,superscriptaddress,showpacs,floatfix]{revtex4}

\usepackage{graphicx}% Include figure files
\usepackage{SIunits}
\usepackage[normal]{subfigure}

\newcommand{\comments}[1]{}

\newcommand{\mrm}[1]{\ensuremath{\mathrm{#1}}}

\begin{document}

\title{Measuring the effective phonon density of states of a quantum dot}

\author{K.~H.~Madsen}\email{khmadsen@nbi.ku.dk}
\affiliation{Niels Bohr Institute, University of Copenhagen, Blegdamsvej 17, DK-2100 Copenhagen, Denmark}
\affiliation{DTU Fotonik, Department of Photonics Engineering, Technical University of Denmark, \O rsteds Plads 343, DK-2800 Kgs.\ Lyngby, Denmark}
\author{P.~Kaer}
\affiliation{DTU Fotonik, Department of Photonics Engineering, Technical University of Denmark, \O rsteds Plads 343, DK-2800 Kgs.\ Lyngby, Denmark}
\author{A.~Kreiner-M\o ller}
\affiliation{DTU Fotonik, Department of Photonics Engineering, Technical University of Denmark, \O rsteds Plads 343, DK-2800 Kgs.\ Lyngby, Denmark}
\author{S.~Stobbe}
\affiliation{Niels Bohr Institute, University of Copenhagen, Blegdamsvej 17, DK-2100 Copenhagen, Denmark}
\author{A.~Nysteen}
\affiliation{DTU Fotonik, Department of Photonics Engineering, Technical University of Denmark, \O rsteds Plads 343, DK-2800 Kgs.\ Lyngby, Denmark}
\author{J.~M\o rk}
\affiliation{DTU Fotonik, Department of Photonics Engineering, Technical University of Denmark, \O rsteds Plads 343, DK-2800 Kgs.\ Lyngby, Denmark}
\author{P.~Lodahl}\email{lodahl@nbi.ku.dk} \homepage{www.quantum-photonics.dk}
\affiliation{Niels Bohr Institute, University of Copenhagen, Blegdamsvej 17, DK-2100 Copenhagen, Denmark}

\date{\today}

\begin{abstract}
We employ detuning-dependent decay-rate measurements of a quantum dot in a photonic-crystal cavity to study the influence of phonon dephasing in a solid-state quantum-electrodynamics experiment. The experimental data agree with a microscopic non-Markovian model accounting for dephasing from longitudinal acoustic phonons, and identifies the reason for the hitherto unexplained difference between non-resonant cavity feeding in different nanocavities. From the comparison between experiment and theory we extract the effective phonon density of states experienced by the quantum dot. This quantity determines all phonon dephasing properties of the system and is found to be described well by a theory of bulk phonons.
\end{abstract}

\pacs{78.67.Hc, 63.20.kd, 42.50.Pq, 03.65.Yz}
%42.50.Pq 	Cavity quantum electrodynamics; micromasers
%78.67.Hc 	Quantum dots
%63.20.kd 	Phonon-electron interactions
%03.65.Yz 	Decoherence; open systems;
%42.50.Ct 	Quantum description of interaction of light and matter; related experiments
\maketitle

Mechanical motion due to thermal vibrations, i.e., phonons, are omnipresent in solid-state systems and inevitably lead to decoherence of quantum superpositon states encoded in the system. Understanding and ultimately engineering such phonon processes may lead to new opportunities for coherent light-matter interaction in an all-solid-state environment. So far, the intrinsic phonon-dephasing mechanisms have not been unravelled, which implies that the full potential of quantum dots (QDs) for on-chip quantum-photonics applications has not yet been explored \cite{Obrien.NaturePhotonics.2009}. For example, it has been predicted that a proper account of the phonon decoherence mechanism is required to achieve highly indistinguishable single photons from QDs~\cite{Kaer.Preprint.ID}. Furthermore, the emerging field of quantum optomechanics explores the ultimate quantum mechanical motion of, e.g., nanomembranes~\cite{Kippenberg.Science.2008}, and generating carriers in semiconductor membranes have proven to enable novel mechanical cooling mechanisms~\cite{WilsonRae.PRL.2004,Usami.NaturePhysics.2012}.

Cavity quantum electrodynamics (QED) experiments in solid-state systems have experienced major breakthroughs within the last decade, including the observation of Purcell enhancement~\cite{Gerard.PRL.1998,Englund.PRL.2005}, strong coupling between a single QD and a photon ~\cite{Yoshie.Nature.2004,Hennessy.Nature.2007}, and non-Markovian dynamics \cite{Madsen.PRL.2011}. A number of surprises have emerged  distinguishing QD-based QED from its atomic counterpart, including the break-down of the point-dipole description of light-matter interaction \cite{Andersen.Nature.2011}, the role of phonon dephasing \cite{Hohenester.PRB.2009,Calic.PRL.2011,Kaer.PRL.2010}, and multiple-charge transitions ~\cite{Winger.PRL.2009,Chauvin.PRB.2009}. In this Letter, we explore the role of environmental fluctuations by employing a QD embedded in a nanocavity as a sensitive probe of the phonon dephasing processes in a photonic crystal (PC) nanomembrane. We compare our experimental data to a microscopic theory for longitudinal acoustic (LA) phonons and extract the effective phonon density of states (DOS) for the QD, which is the key concept describing all aspects of phonon dephasing. Previous work on detuning-dependent decay rates in a PC nanocavity has established the importance of phonon dephasing for the Purcell enhancement~\cite{Hohenester.PRB.2009}. In the present experiment, the dynamics of a single QD enables a direct measurement of the phonon decoherence mechanism, which is possible since the QD is embedded in a coherent cavity. Our work constitutes a significant extension, into the realm of coherent quantum optics, of previous work where proper account for the QD finestructure enabled probing the incoherent local optical density of states (LDOS)~\cite{Wang.PRL.2011}. The applied method is expected to have widespread applications for probing phonon dephasing in advanced nanostructures where the combination of photonic and phononic band gaps \cite{Eichenfield.Nature.2009} ultimately could enable complete coherent control over single-photon emission from QDs.

\begin{figure}%[t!!]
\includegraphics[width=83mm]{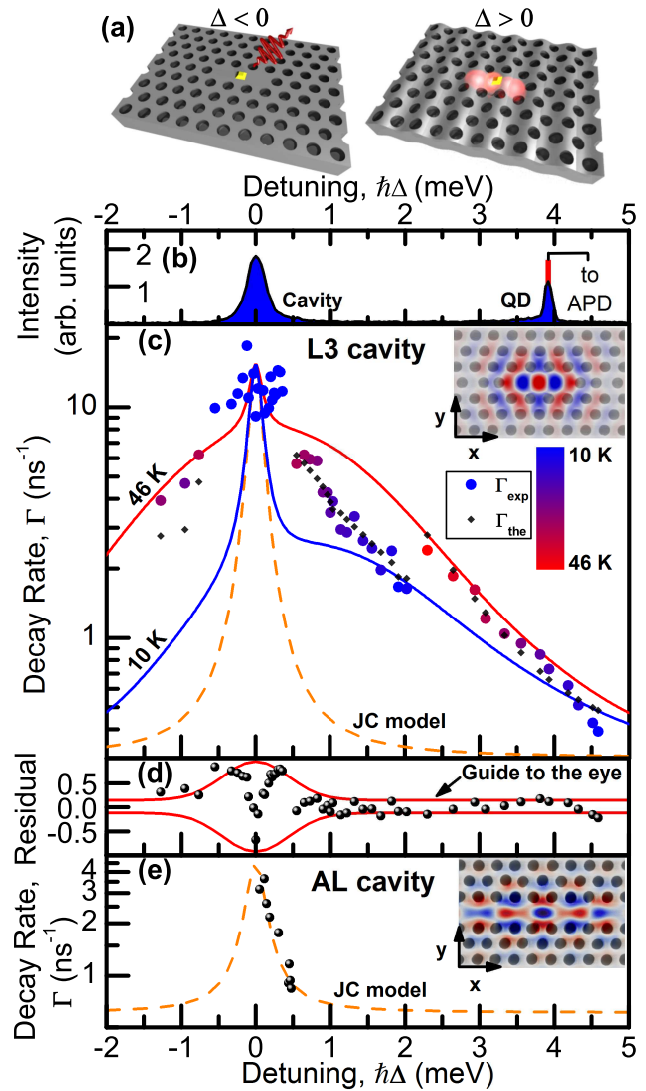}
\caption{(Color online). \textbf{a)} Illustration of physical mechanism behind phonon-enhanced Purcell effect. For negative (positive) detuning, the QD (yellow point) can decay into the cavity by absorbing (emitting) a phonon. At a vanishing temperature no phonon absorption is possible and residual spontaneous emission (red wavepacket) will dominate for negative detuning. \textbf{b)} Emission spectrum with the indication of spectral range detected by APD. \textbf{c)} Measured decay rate $(\Gamma_{exp})$ and calculated rate $(\Gamma_{the})$ at the experimentally used temperature versus detuning. The color bar indicates the corresponding temperature. The comparison to theory is plotted for the largest and smallest temperatures (solid curves). The dotted orange curve is the prediction from the JC model. The applied excitation power is $1.5$ times the saturation power of the QD except for $\Delta<-0.5$ meV and $\Delta>2$ meV, where the power is $0.7$ times the saturation power. \textbf{d)} Residuals $(\Gamma_{exp}-\Gamma_{the})/\Gamma_{exp}$ between experiment and theory. \textbf{e)} Decay rate measurements for an Anderson-localized cavity in a disordered PC waveguide that are well explained by a JC model. Insets in \textbf{c)} and \textbf{e)} show SEM images of the physical systems, where the $y$-components of simulated electric field is overlayed. \label{fig:1}}
\end{figure}

We investigate a GaAs PC membrane with lattice constant $a=240$ nm, hole radius $r=65$ nm, and thickness $154$ nm containing self-assembled InGaAs QDs with a density of $\sim80 \; \micro \metre^{-2}$. An L3 cavity is introduced by leaving out three holes (see Fig.~\ref{fig:1}(c) inset) and shifting the three first holes at each end of the cavity by $0.175a$, $0.025a$, and $0.175a$, respectively~\cite{Akahane.OpticsExpress.2005}. We measure $Q=6690\pm37$ corresponding to a cavity decay rate of $\hbar \kappa = 195 \pm 1 \; \micro$eV by saturating the QDs and recording the cavity linewidth. We employ confocal microscopy and the collected emission is sent to an avalanche photo detector (APD) for time-resolved measurements. Whereas above-band excitation gives rise to emission from many QDs and multiexcitons within each QD, we tune the excitation laser into resonance with higher-order modes of the cavity, thus enabling selective excitation of QDs positioned within the cavity. The fundamental and high-Q mode of the cavity (M1) is observed at $952$ nm, and we use the sixth cavity mode (M6) situated at $850$ nm for excitation \cite{Kaniber.NJP.2009}.

In cavity QED, phonons in the solid-state environment can significantly alter the dynamics of the QD \cite{Kaer.PRL.2010}. Thus, a QD detuned many linewidths away from the cavity resonance can emit photons to the cavity with a Purcell-enhanced rate through the accompanying  emission (absorption) of a phonon for positive (negative) detuning, which is defined as the frequency difference between the QD transition and the cavity mode, $\Delta=\omega_{\mathrm{qd}}-\omega_{\mathrm{ca}}$. At low temperatures, a large detuning asymmetry is expected since the thermal phonon occupancy is low. For example, at $T=10$ K and phonon energies larger than $1$ meV, which are typical values in the present experiment, we estimate $n \le 0.45$ meaning that the probabilities for phonon emission ($\propto n+1$) and absorption ($\propto n$) vary significantly. The phonon-assisted decay is illustrated in Fig.~\ref{fig:1}(a). For positive detuning the QD can decay through the cavity and in this process creates lattice vibrations, while the corresponding absorption process is suppressed for negative detunings and the QD decays mainly by coupling to radiation modes. The phonon-assisted processes are inherently non-Markovian implying that the memory of the phononic reservoir cannot be neglected and the phonon reservoir is "colored", i.e., frequency dependent \cite{Kaer.PRL.2010}. As explained below, this non-Markovian frequency dependence can be probed experimentally by detuning-dependent decay-rate measurements of a single QD in a cavity.

In Fig.~\ref{fig:1}(c) the QD decay rate is plotted as a function of detuning. The fast rate from the biexponential decay curve contains both radiative and non-radiative contributions~\cite{Johansen.PRB.2010}. Temperature control ($10-46$ K) and gas deposition on the sample enable redshifting the QD and cavity frequency respectively, which allows us to control the detuning in the experiment over a large spectral range enabling the extraction of the phonon DOS over a large energy range. We observe Purcell enhancement that is spectrally much broader than expected from the standard Markovian Jaynes-Cummings (JC) model. An example of a spectral measurement is shown in Fig.~\ref{fig:1}(b). We observe the previously reported long-range coupling between QD and cavity by the strong emission from the cavity even for large detunings~\cite{Ates.NaturePhotonics.2009}, however, the spectral domain is not well suited for extracting phonon dephasing processes of single excitons since multiple charge configurations in the QDs have been shown to feed the cavity~\cite{Winger.PRL.2009}. For QDs tuned out of the cavity, see Fig.~\ref{fig:1}(b), the dynamics of single exciton lines under the sole influence of phonon dephasing is considered. For smaller detunings, $\hbar|\Delta|<0.5$ meV, cavity feeding from additional exciton lines influences the decay curves that consequently appear to be multi-exponential, cf. data in Fig.~\ref{fig:1}(c), and we fit these decay curves with triple-exponentials and extract the fast rate that is expected to be dominated by the resonant exciton. In the present experiment we focus mainly on the large-detuning case where phonon-dephasing processes are reliably extracted.

\begin{figure}[tb]
\includegraphics[width=83mm]{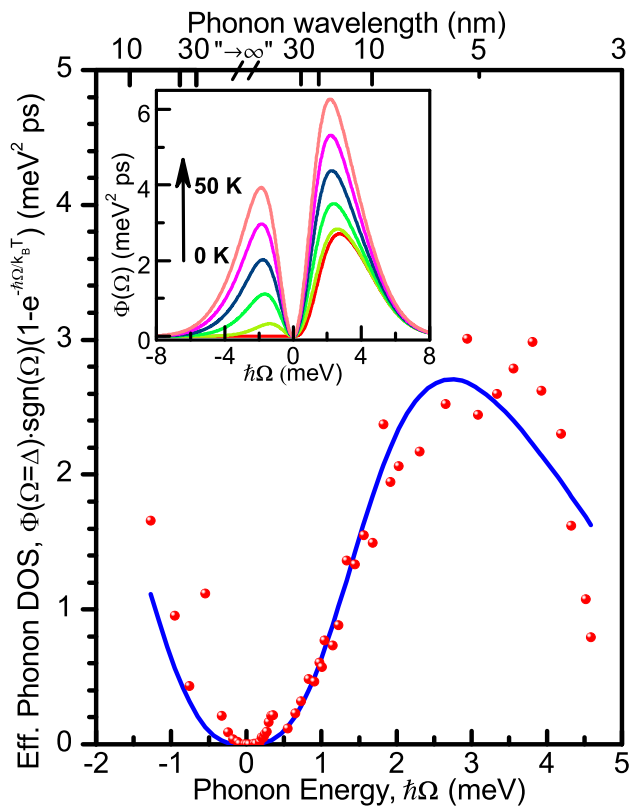}
\caption{(Color online). Effective phonon DOS for a QD versus phonon energy extracted from the data in Fig.~\ref{fig:1}(c). The upper axis indicates the corresponding phonon wavelength.  At large phonon energies (short wavelengths) a cutoff is observed where  the DOS drops. The blue curve shows the corresponding theory assuming LA phonons in a bulk GaAs. The inset shows the role of temperature on the effective phonon density.  \label{fig:3}}
\end{figure}

We describe the QD-cavity coupling with a microscopic model that accounts for the solid-state environment through LA-phonon dephasing of the polariton quasi-particles formed in the cavity. Recently, photon-phonon interaction has been demonstrated in optically pumped L3 PC cavities~\cite{Gavartin.PRL.2011}, but such effects are many orders of magnitude smaller than the other coupling terms in the Hamiltonian for the situation of a single-QD photon source considered here. Memory effects of the phonon bath are taken into account and in the weak-coupling regime the detuning-dependent decay rate of the QD is expressed as~\cite{Kaer.Preparation}
\begin{eqnarray}\label{eq:2}
\Gamma= \gamma+2g^2\frac{\gamma_{\text{tot}}}{\gamma^2_{\text{tot}}+\Delta^2}\left[ 1+\frac{1}{\hbar^2 \gamma_{\text{tot}}} \Phi(\Omega=\Delta)\right],
\end{eqnarray}
where $g$ is the light-matter coupling strength and $\gamma_\mrm{tot}=(\gamma+\kappa)/2$, where $\gamma$ is the decay rate associated with coupling to radiation modes and nonradiative recombination. $\Phi(\Omega=\Delta)$ serves as an  effective phonon density experienced by the QD and evaluated at the frequency, $\Omega$, corresponding to the cavity detuning. The calculated $\Phi(\Omega=\Delta)$ versus detuning is plotted in the inset of Fig.~\ref{fig:3} for increasing temperature. It vanishes for negative detuning at $T=0 \: \mathrm{K}$, since here no phonons are available for absorption, while it is non-zero for positive detuning since the QD can spontaneously emit a phonon enabling the emission of a photon at the cavity frequency. As the temperature increases, the asymmetry gradually levels out since the imbalance between phonon absorption and emission processes disappears. For the experimental decay rate $\Gamma_{exp}$ we also plot the decay rate predicted from theory $\Gamma_{the}$ calculated at the temperature of the particular experiment, for detunings $\hbar|\Delta|>0.5$ meV where a single QD line can be resolved. The excellent agreement between experiment and theory is illustrated by the residuals plotted in Fig.~\ref{fig:1}(d). Eq.~(\ref{eq:2}) constitutes a direct link between the effective phonon density and the QD decay rate. This reveals the interesting insight that by embedding a QD in a cavity the detailed information about dephasing can be extracted from the dynamics of the QD population. The efficient coupling of the QD to the cavity strongly amplifies the weak phonon effects thereby making them measurable and in the absence of the cavity  coherent effects are only revealed from the QD polarization. Indeed increasing the light-matter coupling strength (i.e., $g$) implies that the term containing the phonon density increases relative to the background decay rate $\gamma$ thereby making it visible, as described by Eq. (\ref{eq:2}).

The broadband Purcell enhancement in Fig.~\ref{fig:1}(c) was not observed in previous work on micropillar cavities \cite{Madsen.PRL.2011}, and in order to explain and exploit the origin of these differences we have experimentally studied yet  another type of solid-state cavity system, namely an Anderson-localized (AL) cavity formed in a PC waveguide with scattering imperfections. The sample is displayed in the inset of Fig.~\ref{fig:1}(e). Here, random cavities are generated by randomly perturbing the hole positions in the three rows of holes on each side of the waveguide with a standard deviation of $3\%$ of the lattice parameter $a$. Cavity QED with this sample was previously reported~\cite{Sapienza.Science.2010}. In the AL cavity we observe no broadband Purcell enhancement, and the detuning dependence is well described by the Markovian JC model~\cite{Carmichael.PRA.1989}, cf. Fig.~\ref{fig:1}(e), despite the fact that the measured coupling strength and Q factor of the AL and L3 cavities do not differ significantly. Thus we find $\hbar \kappa_{AL}=230 \pm 12 \; \micro$eV $(Q=5700 \pm 288)$ and $\hbar g_{AL}=13.3 \; \micro$eV  for the AL cavity, which should be compared to $\hbar \kappa=195 \pm 1\; \micro$eV and $\hbar g = 22 \pm 0.7\; \micro$eV for the L3 cavity. As we demonstrate below, a crucial difference between the two cavities stems from the different background decay rates that are obtained from far-detuned decay rates. We record $\hbar \gamma=0.2\;\micro$eV for the L3 cavity and $\hbar \gamma_{AL}=0.4\;\micro$eV for the AL cavity.

\begin{figure}[tb]
\includegraphics[width=83mm]{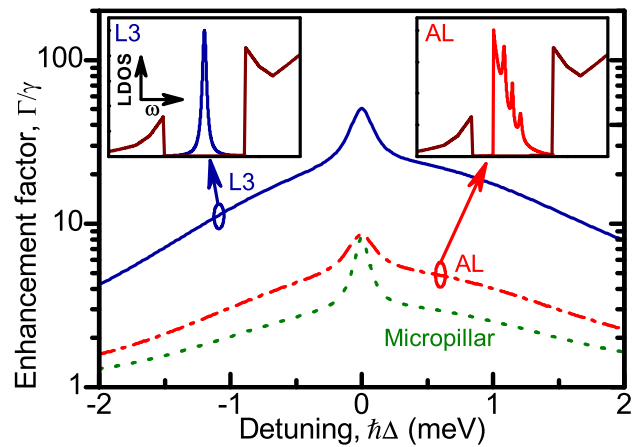}
\caption{(Color online). Calculation of the detuning dependence of the enhancement factor for three different cavity QED systems: the L3 PC cavity, an AL cavity, and a micropillar cavity at $T=30$ K. For the first two systems the parameters extracted from the present experiment are used while the latter corresponds to the experiment of Ref. \onlinecite{Madsen.PRL.2011}. Insets show sketches of the optical LDOS for the L3 cavity and AL cavity. \label{fig:2}}
\end{figure}

The difference between AL and micropillar cavities on the one hand and L3 PC cavities on the other can be explained from Eq.~(\ref{eq:2}). In Fig.~\ref{fig:2} we show the Purcell enhancement factor $\Gamma/\gamma$ for three different cavities calculated using Eq.~(\ref{eq:2}) with identical effective phonon densities and all additional parameters determined from experimental data. We find that $\gamma$ plays a decisive role in determining the visibility of the phonon influence on the decay dynamics, and it varies significantly  for the different cavity geometries.  The insets in Fig.~\ref{fig:2} show a sketch of the LDOS for the L3 and AL cavities highlighting that while L3 PC cavities appear in the band gap where the background decay rate is strongly suppressed~\cite{Wang.PRL.2011}, AL cavities appear as random resonances on top of a background LDOS representing the waveguide mode \cite{Sapienza.Science.2010} giving rise to an emission channel. Similarly in micropillar cavities the coupling to radiation modes is not strongly suppressed. In more quantitative terms two requirements need to be fulfilled in order to see broadband Purcell enhancement: $\frac{\Phi(\Delta)}{\hbar^2 \gamma_{tot}}\geq 1$ and $\frac{2g^2}{\hbar^2 \gamma} \frac{\Phi(\Delta)}{\Delta^2}\geq1$. The former weak condition is for a cavity with $Q=6690$ fulfilled for detunings above $0.45$ meV. The latter and stronger condition  can be evaluated to $ \frac{2g^2}{\gamma} \times (1.2 \;\mathrm{ps}) = 4.47\geq1$ for a typical detuning of $2$ meV. This value is $5.5$ times larger for the L3 compared to the AL cavity.

Fig.~\ref{fig:3} shows the measured effective phonon DOS extracted from the data in Fig.~\ref{fig:1}(c).
In order to compensate for the effect of temperature through the phonon occupation, we plot the temperature-independent quantity $\Phi(\Omega=\Delta) \cdot \mathrm{sgn}(\Delta)(1-e^{-\hbar \Omega/k_\mathrm{B}T})$, which is the effective phonon DOS experienced by the QD. The observed energy dependence of the effective phonon DOS is a direct signature of the non-Markovianity of the phonon reservoir. The phonon DOS is low for small phonon energies and grows rapidly with energy, reaching a maximum at about $3$ meV. This energy cutoff originates from the phonon wavelength (of about $5$ nm) becoming comparable in size to the wavefunction of the exciton confined in the QD~\cite{Ramsay.PRL.2010}. Thus, tailoring the QD size may be employed for influencing phonon-dephasing processes. We stress that the applied method for extracting the phonon DOS is general and may be exploited in the future in more advanced nanostructures combining photonic and phononic band gaps, thus potentially enabling complete coherent control over the single-photon emission from the QDs, which will be very valuable for all-solid-state quantum information processing.

We finally compare our experimental data of the effective phonon DOS to calculations assuming bulk phonons in GaAs, cf. Fig.~\ref{fig:3}. Very good agreement between experiment and theory is observed, in particular for energies larger than $0.5$ meV where the effects of multiexciton processes or other QDs are negligible. The remarkable success of the bulk-phonon theory excludes effects of localization of phonon modes in the cavity. Phonon localization in an L3 PC cavity has been reported at energies up to $4.1$ $\micro$eV~\cite{Gavartin.PRL.2011}, while here typical phonon energies required for the photon-assisted recombination are above $0.5$ meV. The corresponding phonon wavelength is below $42$ nm, which is much smaller than the dimensions of the cavity, explaining the success of the bulk-phonon theory. We note that localized phonons could play a role for the minor deviations from theory observed at small phonon energies. In the theory, we expand the bulk phonons in plane waves, and both the excited and ground state electron wavefunctions in the QD are assumed to be gaussian in all three dimensions and circular symmetric in the plane orthogonal to the growth direction. We successfully model the data with the following realistic widths of the electron wavefunctions; $l_{\mathrm{e,xy}}=3.4$ nm, $l_{\mathrm{e,z}}=1.4$ nm, $l_{\mathrm{g,xy}}=3.9$ nm, and $l_{\mathrm{g,z}}=2.3$ nm, where $\mathrm{e}$ and $\mathrm{g}$ denote the excited and ground state, respectively.
These sizes also determine the oscillator strength of the QD to $15.5$, in very good agreement with previous measurements~\cite{Johansen.PRB.2008}. Remarkably, the only additional free parameter is an overall scaling factor of the curve of $5.56$, which corresponds to dividing the speed of sound by $1.41$~\cite{ReD.Expression}. This is attributed to the anisotropy of sound velocity in GaAs, which is not explicitly accounted for in the theory. We stress the importance of accounting for the microscopic non-Markovian dephasing processes: for large detunings the quantity $\Phi(\Delta\gg0)\hbar^{-2}$ enters in the theory as an effective pure-dephasing rate, but it is typically $3$ orders of magnitude larger than dephasing rates extracted when interpreting experiments with a Markovian model~\cite{Laurent.APL.2005}.

In conclusion, we have observed broadband Purcell enhancement for a QD in a PC cavity.  Our data are very well explained by a microscopic non-Markovian LA phonon theory. We record for the first time the effective phonon DOS of a QD by employing a PC nanocavity to increase sensitivity. Remarkably, our measurements are well described by a bulk phonon theory despite the inhomogeneity of PC cavities. This work is an essential step towards understanding and potentially controlling and engineering the coherence properties of QD-based QED systems, which is required when applying such an all-solid-state platform for quantum-information processing.

We gratefully acknowledge  financial support from the Villum Foundation,
the Danish Council for Independent Research (Natural Sciences and Technology and Production
Sciences) and the European Research Council (ERC consolidator grant "ALLQUANTUM").


\begin{thebibliography}{16}
\bibitem{ReD.Expression}
$\Phi(\Omega)=\frac{\hbar}{4 \pi d c^5_l} \frac{\Omega^3}{1-e^{-\hbar \Omega/k_\mathrm{B}T}}
\int^1_0 du(D_\mathrm{e}e^{\tilde{\Omega}^2_{e,\mathrm{xy}}(u^2-1)-\tilde{\Omega}^2_{e,\mathrm{z}}u^2} -
D_\mathrm{g} e^{\tilde{\Omega}^2_{g,\mathrm{xy}}(u^2-1)-\tilde{\Omega}^2_{g,\mathrm{z}}u^2})^2$,
where $\tilde{\Omega}_{\nu,\mu} = \frac{\Omega l_\mathrm{\nu,\mu}}{2 c_l}$, and $\nu=\{\mathrm{e,g}\}$ and $\mu=\{\mathrm{xy,z}\}$. $D_{\mathrm{e}}=-14.6 ~ \mathrm{eV}$ and $D_{\mathrm{g}}=-4.8 ~ \mathrm{eV}$ are the deformation potentials, $c_l=5110 ~ \frac{\mathrm{m}}{\mathrm{s}}$ is the longitudinal speed of sound averaged over the three different crystal axes, and $d=5370~\frac{\mathrm{kg}}{\mathrm{m}^3}$ is the density of GaAs.
\end{thebibliography}
\end{document}